\documentclass[a4paper]{jpconf}
\usepackage{graphicx}
\begin{document}
\title{Magnetic incommensurability and fluctuating charge density waves in the repulsive Hubbard model}

\author{A Sherman$^1$ and M Schreiber$^2$}

\address{$^1$ Institute of Physics, University of Tartu, Riia 142, EE-51014 Tartu, Estonia}
\address{$^2$ Institut f\"ur Physik, Technische Universit\"at, D-09107
Chemnitz, Germany}

\ead{alexei@fi.tartu.ee}

\begin{abstract}
Magnetic and charge susceptibilities of the two-dimensional repulsive Hubbard model are investigated applying a strong coupling diagram technique in which the expansion in powers of the hopping constants is used. For small lattices and high temperatures results are in agreement with Monte Carlo simulations. With the departure from half-filling $x$ the low-frequency magnetic susceptibility becomes incommensurate and the incommensurability parameter grows with $x$. The incommensurability, its dependence on frequency and on $x$ resemble experimental results in lanthanum cuprates. Also for finite $x$ sharp maxima appear in the static charge susceptibility. The maxima are finite which points to the absence of the long-range charge ordering (static stripes). However, for $x\approx 0.12$ the maxima are located near the momenta $(0,\pm\pi/2)$, $(\pm\pi/2,0)$. In this case an interaction of carriers with tetragonal distortions can stabilize stripes with the wavelength of four lattice spacings, as observed in the low-temperature tetragonal phase of cuprates. As follows from the obtained results, the magnetic incommensurability is not a consequence of the stripes.
\end{abstract}

The Hubbard model is thought to be appropriate to describe the main features of electron correlations in narrow energy bands, leading to collective effects such as magnetism and metal-insulator transition. It has often been used to describe real materials exhibiting these phenomena. In more than one dimension, the model is not exactly solvable and a variety of numerical and analytical approximate methods was used for its study. In the case of strong electron correlations inherent in cuprate perovskites, when the ratio of the hopping constants to the Hubbard repulsion is a small parameter, the strong-coupling diagram technique \cite{Vladimir,Sherman06} can be applied for the investigation of the model. In this technique, Green's functions are calculated using expansions in powers of the hopping constants. The terms of these expansions are expressed by means of site cumulants of electron creation and annihilation operators.

In this article we use this approach for calculating magnetic and charge susceptibilities of the 2D repulsive Hubbard model which presumably describes the Cu-O plane of cuprate perovskites. Our aim is to investigate possible mechanisms of the incommensurate magnetic response \cite{Mason} and the stripe formation \cite{Tranquada} observed in these crystals. The susceptibilities are connected with the spin and charge Green's functions
\begin{equation}\label{Green}
D({\bf l'\tau,l\tau})=\langle{\cal T}s^\sigma_{\bf l'}(\tau') s^{-\sigma}_{\bf l}(\tau)\rangle, \quad B({\bf l'\tau,l\tau})=\langle{\cal T}\delta n_{\bf l'}(\tau')\delta n_{\bf l}(\tau)\rangle,
\end{equation}
where the angular brackets denote the statistical averaging with the Hubbard hamiltonian which also determines the time evolution of operators in Eq.~(\ref{Green}), ${\cal T}$ is the chronological operator, the spin operators $s^\sigma_{\bf l}=a^\dagger_{\bf l\sigma}a_{\bf l,-\sigma}$, $a^\dagger_{\bf l\sigma}$ and $a_{\bf l\sigma}$ are the electron creation and annihilation operators on the site {\bf l} of the square lattice with the spin projection $\sigma$, $\delta n_{\bf l}=n_{\bf l}-\langle n_{\bf l}\rangle$ with $n_{\bf l}=\sum_\sigma a^\dagger_{\bf l\sigma}a_{\bf l\sigma}$.

Using the strong-coupling diagram technique one can convince oneself that equations for $D$ and $B$ look similar \cite{Sherman07}. For the function $B$ they read
\begin{eqnarray}
&&B(q)=-\frac{T}{N}\sum_{p_1}G(p_1)G(q+p_1)
 +\left(\frac{T}{N}\right)^2\sum_{p_1p_2}\Pi(p_1)\Pi(p_2)\Pi(q+p_1)
 \Pi(q+p_2)\nonumber\\
&&\quad\quad\times\Lambda(p_1,q+p_1,q+p_2,p_2), \nonumber\\[-1ex]
&&\label{eqforB}\\[-1ex]
&&\Lambda(p_1,q+p_1,q+p_2,p_2)=\lambda(p_1,q+p_1,q+p_2,p_2)
 -\frac{T}{N}\sum_{p_3}\lambda(p_1,q+p_1,q+p_3,p_3)\nonumber\\
&&\quad\quad\times\Theta(p_3) \Theta(q+p_3) \Lambda(p_3,q+p_3,q+p_2,p_2). \nonumber
\end{eqnarray}
Here the combined indices $q=({\bf k},i\omega_\nu)$ and $p_j=({\bf
k}_j,i\omega_{n_j})$ were introduced, $\omega_\nu=2\nu\pi T$ and $\omega_n=(2n+1)\pi T$ are the boson and fermion Matsubara frequencies with the temperature $T$, ${\bf k}$ is the wave vector, $G(p)=\langle\langle a_{\bf k\sigma}|a^\dagger_{\bf k\sigma}\rangle\rangle$ is the electron Green's function, $\Pi(p)=1+t_{\bf k}G(p)$, $t_{\bf k}$ is the Fourier transform of the hopping constants, $\Theta(p)=t_{\bf k} \Pi(p)$ is the renormalized hopping, $N$ is the number of sites, $\Lambda(p_1,p+p_1,p+p_2,p_2)$ is the sum of all four-leg diagrams and $\lambda(p_1,p+p_1,p+p_2,p_2)$ is its irreducible subset. With explicitly specified spin indices of the external lines, this latter notation reads $\lambda=\frac{1}{2}\sum_{\sigma' \sigma}\lambda(\sigma',\sigma',\sigma,\sigma)$. The difference between Eq.~(\ref{eqforB}) and the equations for $D$ is in the spin indices of this quantity. In the latter case $\lambda(\uparrow,\downarrow, \downarrow,\uparrow)$ enters into the equations.

In the calculations we approximated quantities $\Pi(p)$ and $\Theta(p)$ by 1 and $t_{\bf k}$, respectively, and used the lowest-order irreducible four-leg diagrams -- the second-order cumulants -- for $\lambda$ in Eq.~(\ref{eqforB}).
\begin{figure}[h]
\includegraphics[width=17pc]{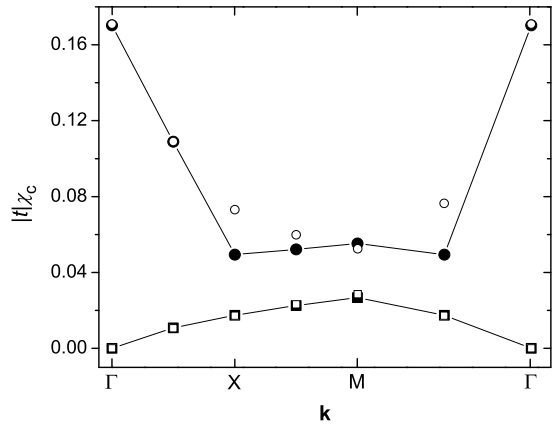}\hspace{2pc}%
\begin{minipage}[b]{18.7pc}\caption{\label{Fig1}The static charge susceptibility in a 4$\times$4 lattice for $t=-U/8$ and $T=0.125|t|$. The susceptibility is plotted along the triangular contour in the Brillouin zone. The corners of the contour are given by the momenta ${\bf k}=(0,0)$ ($\Gamma$), $(\pi,0)$ ($X$), and $(\pi,\pi)$ ($M$). Open symbols are results of Monte Carlo simulations \protect\cite{Bickers} for $\bar{n}=1$ (squares) and $\bar{n}=0.95$ (circles). Filled symbols are our results for the same electron fillings.}
\end{minipage}
\end{figure}
A comparison of the obtained results with Monte Carlo simulations is shown in Fig.~\ref{Fig1}. In this figure, $t$ is the nearest-neighbor hopping constant, $U$ is the Hubbard repulsion and $\bar{n}=\langle n_{\bf l}\rangle$ is the electron filling.

\begin{figure}[t]
\centerline{\includegraphics[height=17pc]{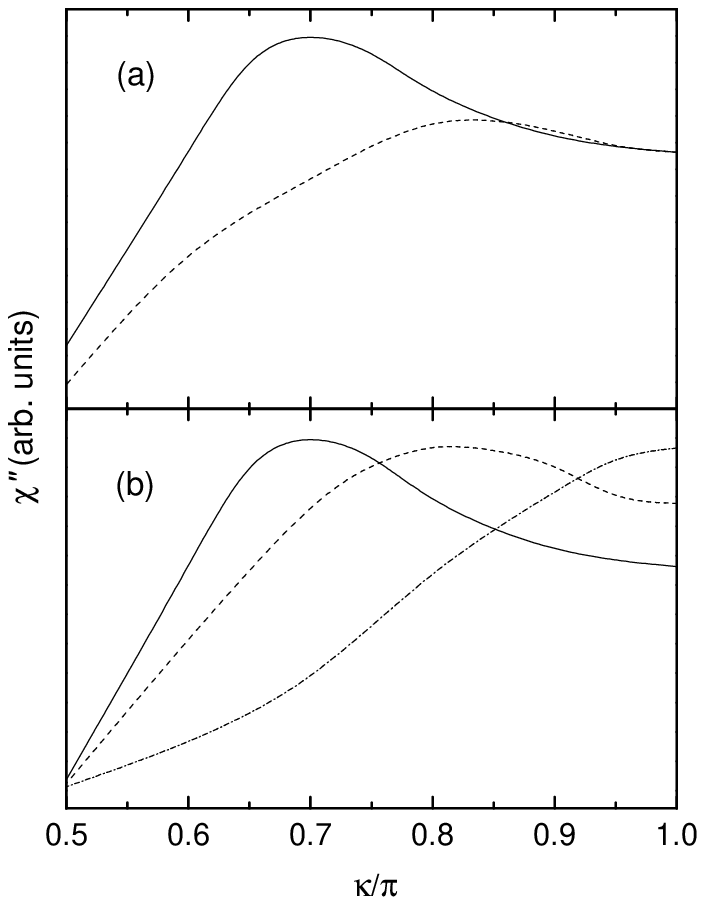}\hspace{2pc}\includegraphics[height=17pc]{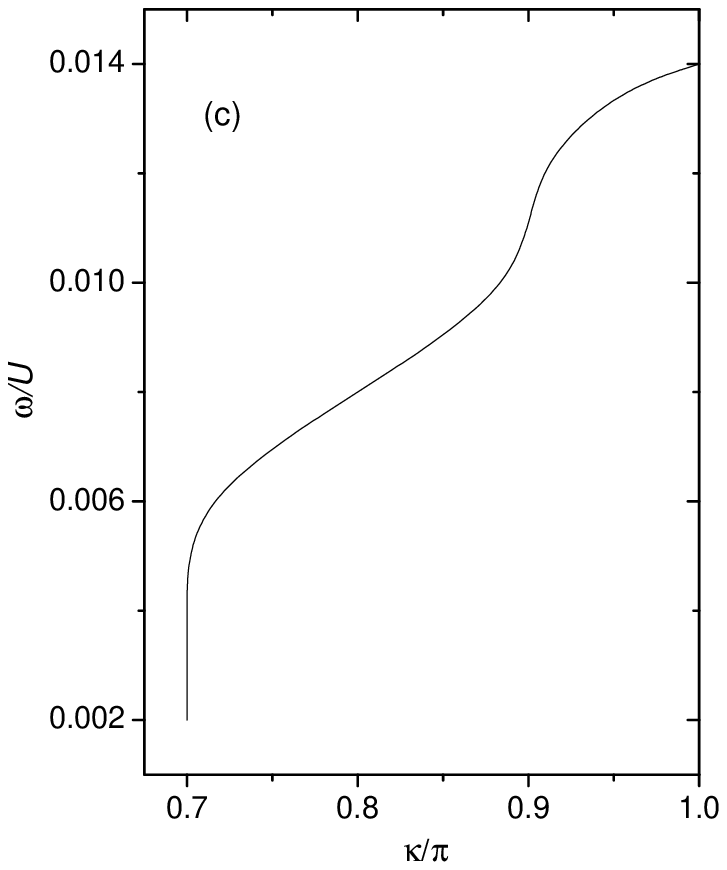}}%
\caption{\label{Fig2}(a) The momentum dependence of $\chi''({\bf k}\omega)$ along the edge [solid line, ${\bf k}=(\pi,\kappa)$] and diagonal [dashed line, ${\bf k}=(\kappa,\kappa)$] of the Brillouin zone for $t=-0.11U$, $\omega=0.002U$ and $\bar{n}\approx 0.88$). (b) The momentum dependence of $\chi''({\bf k}\omega)$ along the zone edge for $\bar{n}\approx 0.88$ (solid line), $\bar{n}\approx 0.94$ (dashed line), and $\bar{n}=1$ (dash-dotted line). $t=-0.11U$ and $\omega=0.002U$. (c) The dispersion of maxima in $\chi''({\bf k}\omega)$ along the zone edge for $t=-0.11U$ and $\bar{n}\approx 0.88$.}
\end{figure}
For small real frequencies $\omega$ the momentum dependence of the imaginary part of the magnetic susceptibility $\chi''({\bf k}\omega)={\rm Im}D({\bf k}\omega)$ calculated in a 100$\times$100 lattice is shown in Fig.~\ref{Fig2}. At half-filling, $\bar{n}=1$, for the considered temperature $T=0.05U$ the system is in the state with the commensurate short-range order [the dash-dotted curve in Fig.~\ref{Fig2}(b)]. With departure from half-filling, the maximum of $\chi''({\bf k}\omega)$ shifts from $(\pi,\pi)$ and the susceptibility becomes incommensurate. For $x=|1-\bar{n}|>0.06$ the susceptibility is peaked at momenta $(\pi,\pi\pm\delta)$, $(\pi\pm\delta,\pi)$ [see Fig.~\ref{Fig2}(a)] and the incommensurability parameter $\delta$ grows with $x$ [see Fig.~\ref{Fig2}(b)]. For a fixed $\bar{n}$, $\delta$ decreases with increasing $\omega$ and, at some frequency $\omega_r$, the incommensurability disappears and $\chi''({\bf k}\omega)$ appears to be peaked at $(\pi,\pi)$ [see Fig.~\ref{Fig2}(c)]. The same behavior of the magnetic susceptibility at low frequencies was observed in lanthanum cuprates \cite{Mason}. In Fig.~\ref{Fig2}, the values of $\delta$ are close to those measured experimentally for the same $x$. Parameters of this figure were chosen so that for $\bar{n}\approx 0.88$ the value of $\omega_r$ is close to the experimental value 50 meV. For the superexchange constant $J=4t^2/U=0.15$~eV we find $\omega_r=44$~meV from Fig.~\ref{Fig2}.

\begin{figure}[t]
\centerline{\includegraphics[height=14pc]{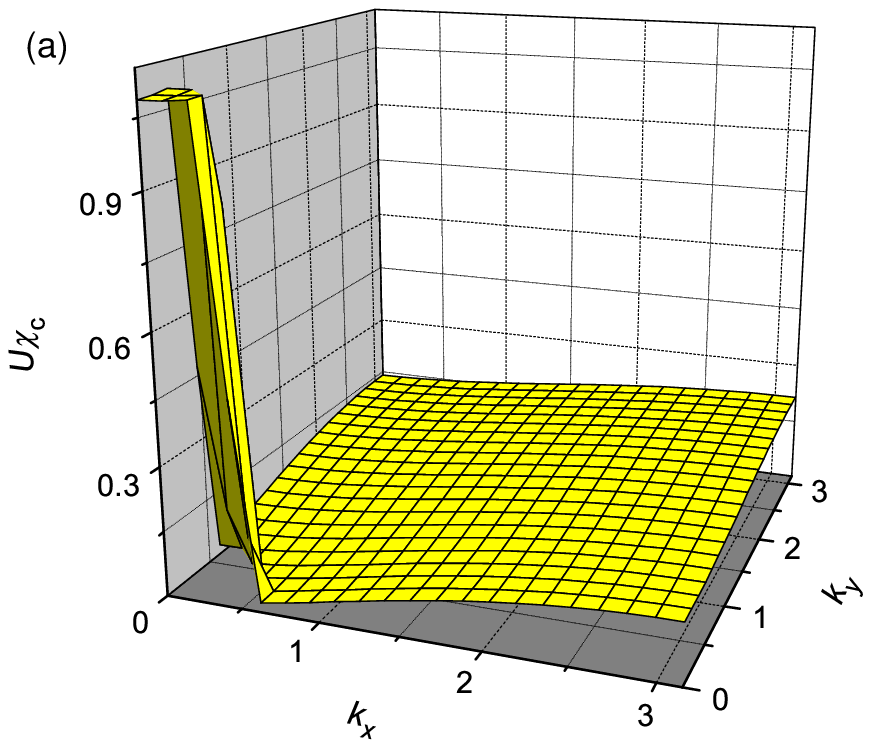}\hspace{1pc}\includegraphics[height=14pc]{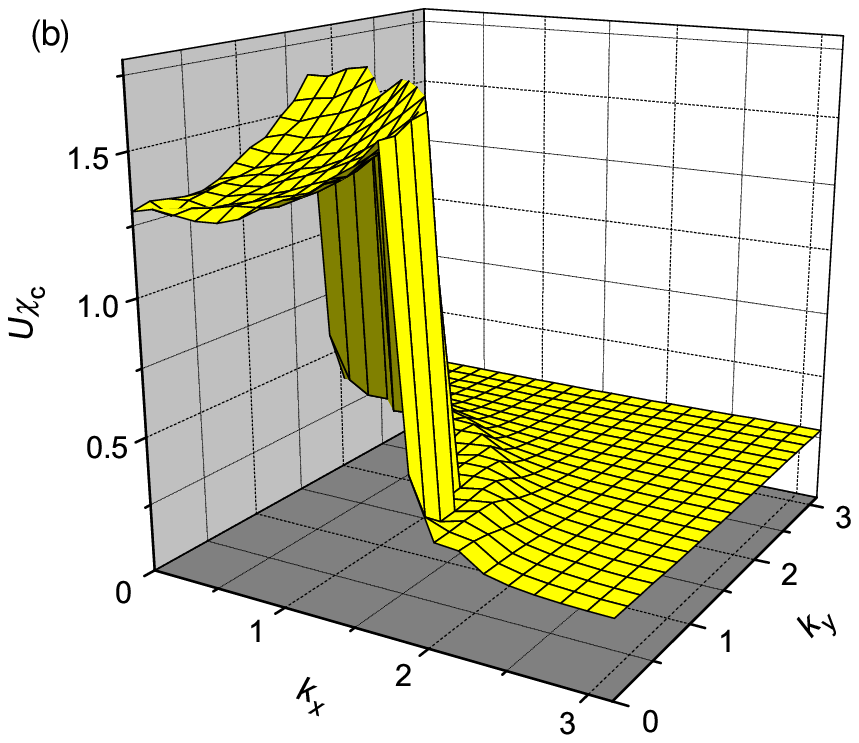}}%
\caption{\label{Fig3}The static charge susceptibility in a 40$\times$40 lattice for $t=-U/8$ and $T=0.001U$. The susceptibility is shown in the first quadrant of the Brillouin zone for the electron fillings $\bar{n}=0.97$ (a) and 0.88 (b).}
\end{figure}

The momentum dependence of the static charge susceptibility $\chi_c({\bf k})=B({\bf k},\nu=0)$ is shown in Fig.~\ref{Fig3} for two electron fillings. At half-filling the susceptibility is small and its dependence on momentum is weak. Immediately after the Fermi level crosses one of the Hubbard subbands, which leads to departure from half-filling, a sharp peak appears in $\chi_c({\bf k})$ near the $\Gamma$ point [see Fig.~\ref{Fig3}(a)]. With increasing $x$ the susceptibility grows and the peak transforms to a ridge around $\Gamma$ [see Fig.~\ref{Fig3}(b)]. Notice that in the entire considered range $0\leq x \leq 0.2$ the static charge susceptibility remains finite. This means that a long-range or stripe ordering of charges {\em does not occur} in the Hubbard model for the considered range of parameters corresponding to cuprate perovskites. The variation of the ratio $U/t$ in this range and the inclusion of the hopping to more distant sites do not change this conclusion.

This result indicates that the above-discussed magnetic incommensurability is not connected with charge stripes and is a consequence of strong electron correlations.

Although the above results do not support a purely electronic mechanism of long-range charge ordering, they give some insight into the way in which phonons can stabilize stripes. An essential role of certain CuO$_6$ octahedra tilts in such stabilization follows from the fact that static stripes were observed only in the low-temperature tetragonal (LTT) phase of lanthanum cuprates \cite{Tranquada}. It is known that such tilts are strongly coupled to the carriers \cite{Pickett}. This interaction leads to the following term in the adiabatic potential:
\begin{equation}\label{apot}
\Delta E=-\frac{1}{2}\sum_{\bf k}\chi_c({\bf k})\phi_{\bf k}^2,
\end{equation}
where the function $\phi_{\bf k}$ contains only even powers of tilt coordinates. As follows from Eq.~(\ref{apot}), finite values of the vibration coordinates give an energy gain and the larger the value of the susceptibility is, the larger energy gain can be achieved. As seen from Fig.~\ref{Fig3}(b), for $x\approx 0.12$ the susceptibility maxima are located on the axes of the Brillouin zone approximately halfway between its center and its boundary. Such maxima give the lowest energy gain for the charge density wave with the wavelength equal to four lattice spacings, as observed in the LTT phase of lanthanum cuprates \cite{Tranquada}. In contrast to other electron fillings such a wave is commensurate with the lattice which is essential for its stability.

In summary, the magnetic and charge susceptibilities of the 2D repulsive Hubbard model were calculated using the strong-coupling diagram technique. It was found that with departure from half-filling the low-frequency magnetic susceptibility becomes incommensurate. This incommensurability, its dependence on frequency and on electron filling are similar to experimental results in lanthanum cuprates. In the range of parameters corresponding to cuprate perovskites the maxima of the static charge susceptibility are finite which points to the absence of the long-range charge ordering (static stripes) in the Hubbard model. However, the shape of the obtained susceptibility for $x\approx 0.12$ suggests that an interaction of carriers with tetragonal distortions can stabilize stripes with the wavelength of four lattice spacings, as observed in the low-temperature tetragonal phase of cuprates. From the obtained results it follows that the magnetic incommensurability is not a consequence of the stripes.

This work was partially supported by the ETF grant No.~6918.

\end{document}